\documentclass[aps,showpacs,preprint]{revtex4}
\usepackage{hyperref}
\usepackage{graphicx}
\usepackage{color}
\begin{document}
\title{Quantum Data Bus in Dipolar Coupled Nuclear Spin Qubits}
\author{
Jingfu Zhang,$^1$ 
Michael Ditty,$^1$  Daniel Burgarth,$^2$ Colm A. Ryan,$^1$ C. M. Chandrashekar,$^{1,3}$ Martin Laforest,$^1$
Osama Moussa,$^1$ Jonathan Baugh,$^1$ and Raymond Laflamme$^{1,3}$ \\
\it {$^1$Institute for Quantum Computing and Department of
Physics,
University of Waterloo, Waterloo, Ontario, Canada N2L 3G1\\
$^2$IMS and QOLS, Imperial College, London SW7 2BK, UK\\
$^3$Perimeter Institute for Theoretical Physics, Waterloo,
Ontario, N2J 2W9, Canada}}

\date{\today}

\begin{abstract}
  We implement an iterative quantum state transfer exploiting the natural
dipolar couplings in a spin chain of a liquid crystal NMR system.
During each iteration a finite part of the amplitude of the state is
transferred and by applying an external operation on only the last
two spins the transferred state is made to accumulate on the spin at
the end point. The transfer fidelity reaches one asymptotically
through increasing the number of iterations. We also implement the
inverted version of the scheme which can transfer an arbitrary state
from the end point to any other position of the chain and entangle
any pair of spins in the chain, acting as a full quantum data bus.
\end{abstract}

\maketitle
\section{Introduction}
   In quantum computation and quantum
communication the transfer of an arbitrary quantum state
 from one qubit to another is a
fundamental element. The most obvious method to implement quantum
state transfer (QST) on an array of qubits is based on a sequence
of SWAP gates for neighboring spins. In spin qubit systems the
SWAP gate (up to a known phase factor) can be implemented through
the evolution of the dipolar coupling between the neighboring
spins for $1/2D$ time by decoupling the other spins, where $D$
denotes the dipolar coupling strength. In experiments, however,
the required decoupling operations are hard to implement if the
spins cannot be individually addressed by spectral selectivity,
e.g., in large-size solid-state NMR systems.
 This makes the direct implementation of such gates in a large spin system
 challenging.

To overcome this problem, schemes based on ''always on'' spin
systems were proposed \cite{Bose03,thesis}. The state can be
transferred with unit fidelity in engineered spin chains or
networks with $XY$ interactions \cite{PST}. However, the required
fine-tuned $XY$ couplings are not found in natural spin systems
\cite{CoryPRL07}. In other schemes based on spin chains with
Heisenberg interactions \cite{Bose03,Feldman} or with a
double-quantum Hamiltonian \cite{CoryPRL07}, the fidelity of the
QST cannot approach unity in scalable systems.

 The above limitations can be relaxed significantly by applying
gate operations to receive and store the transferred state
\cite{BurgarthPRA07,zhang07}. The gates are only applied to two
spins at one end of a spin chain. In this paper, we experimentally
implement the QST in a liquid crystal NMR system based on this
scheme. Opposed to previous experimental implementations
\cite{zhang07} where the required $XY$ interactions were engineered
by radio-frequency pulses and scalar couplings, the dipolar
couplings exist naturally in the system and are directly exploited
for the QST. The dipolar couplings are much stronger (up to 2-3
orders of magnitude) than the scalar couplings and therefore can
significantly speed up the implementation of the logical gates for
quantum information processing \cite{Mahesh03}. The transfer with
high fidelity is achieved in an iterative manner. Each iteration
transfers a finite part of the input amplitude to the target spin at
the end of the chain. The fidelity of the transfer asymptotically
approaches unity by increasing the number of iterations. We also
experimentally demonstrate the time-inverted version of
\cite{BurgarthPRA07}. Through this, a full \emph{quantum data bus}
is implemented, where arbitrary unknown quantum states can be
steered to an \emph{any} position of the chain. This is also useful
for the selective excitation of one spin, which is addressed by the
two-spin gates, rather than by its individual properties, e.g.
chemical shift in NMR. As opposed to previous schemes \cite{global},
global control is not required. Surprisingly the reversal operation
can also be used to \emph{entangle} any pair of spins in the chain
by operations at its end only. We demonstrate the entangling
operation in the qubits at the end points of the chain.

 The QST and its reversal operations mean that
 the chain
is really used as a wire with an input, an output, and no gates in
the middle; the many-body Hamiltonian of the chain is responsible
for the transport. Only two spins at the end are required to
address. The fidelity of transfer converges exponentially fast to
unity with respect to the number of iterations.
The required number of iterations to achieve a good fidelity (e.g.
larger than $0.999$) scales roughly linearly with the system size
\cite{BurgarthPRA07}. Moreover this method is stable when the
engineered Hamiltonian in implementation deviates the required
Hamiltonian \cite{EPJ07}. Hence our method scales favorably with
the size of the spin chains and suitable for large-size systems,
such as solid or liquid crystal NMR systems, where the differences
of the chemical shifts are too small to address all the spins
individually.

\section{Iterative transfer algorithm in a spin chain} 
Our first goal is to transfer the state
$\alpha|0\rangle+\beta|1\rangle$ from spins $j$ to $N$ in a $N$-
spin chain. The Hamiltonian for spins $1$ to $N-1$ is represented
as
\begin{equation}\label{Ham123}
    H=\frac{1}{2}\pi
     \sum_{j,k=1;k>j}^{N-1}D_{jk}(2\sigma_{z}^{j}\sigma_{z}^{k}-
    \sigma_{x}^{j}\sigma_{x}^{k}-\sigma_{y}^{j}\sigma_{y}^{k}),
\end{equation}
where $\sigma_{x}^{j}$, $\sigma_{y}^{j}$ and $\sigma_{z}^{j}$ denote
the Pauli matrices with $j$ indicating the affected spin. Noting
that $H$ preserves the total number of excited spins
\cite{PST,Bose03}, we have
\begin{equation}\label{Ut0}
U_{\tau}| \mathbf{0}\rangle=e^{i\theta}|\mathbf{0}\rangle
\end{equation}
\begin{equation}\label{Utj}
U_{\tau}|\mathbf{j}\rangle=\sum_{k=1}^{N-1}a_{k}|\mathbf{k}\rangle
\end{equation}
where $U_{\tau}=e^{-i\tau H}$ and $e^{i\theta}$ is the $(1,1)$
matrix element of $U_{\tau}$. The state $| \mathbf{0}\rangle$
denotes all spins pointing up, and $|\mathbf{j}\rangle$ denotes
all spins up except the spin $j$ pointing down.

   The main operation is the two-spin gate applied only on spins
$N-1$ and $N$, and in iteration $n$ the gate is denoted as
\begin{eqnarray}\label{endgate}
   W(c_n,d_n)=  I_{1,2,...N-2}\bigotimes \left (\begin{array}{cccc}
                   1 & 0 & 0 & 0 \\
                   0 & d_{n}^{*} & c_{n}^{*} & 0 \\
                   0 & -c_{n} & d_{n} & 0 \\
                   0 & 0 & 0 & 1
                 \end{array}
    \right )_{N-1,N}
\end{eqnarray}
where $I_{1,...,N-2}$ denotes the unit operator for spins $1$ to
$N-2$. The basis order for spins $N-1$ and $N$ is $|00\rangle,
|01\rangle, |10\rangle$ and $|11\rangle$. Noting
$|c_n|^2+|d_n|^2=1$, one finds
\begin{equation}\label{accumulate}
    W(c_n,d_n)(c_n|\mathbf{N-1}\rangle+d_n|\mathbf{N}\rangle)=|\mathbf{N}\rangle.
\end{equation}

\begin{figure}
\includegraphics[width=4.5in]{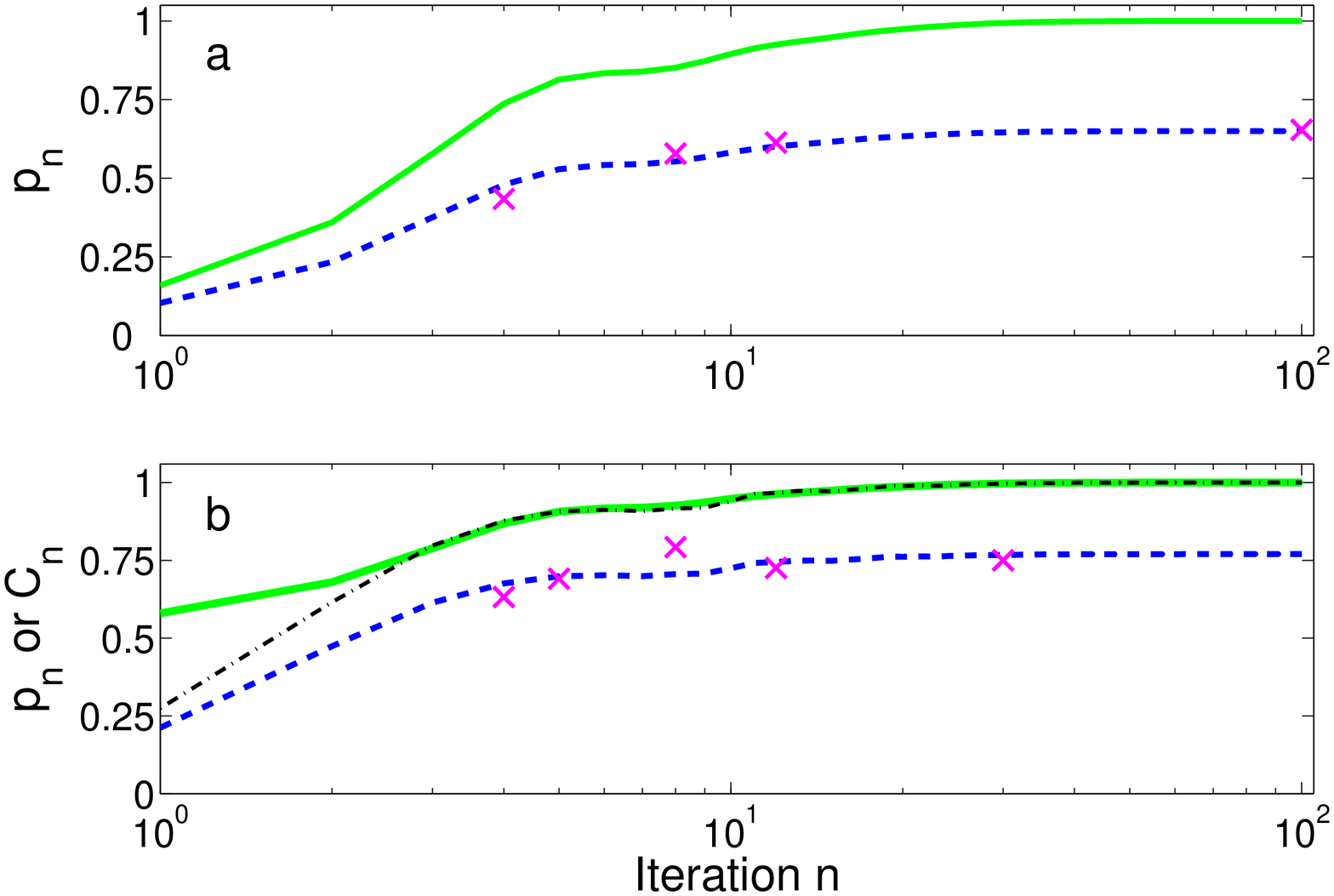}
\caption{(Color online) The numerical simulation (solid) and
experimental results (data marked by ''$\times$'') for the
probability $p_n$ of the QST as a function of iterations in the
four spin system used in experiments (see text for dipolar
couplings) for transferring a state from spins $1$ to $4$ (a) and
entangling the two spins (b) when $\tau=2.1$ ms. In figure (b)
$p_{n}$ can be approximated as the observable coherence $C_n$
[dot-dashed, see Eq. (\ref{Coherence})] where $|C_n-p_n|\leq
0.0175$ when $n>2$. The experimental data can be fitted as
$0.65p_n$ and $0.77C_n$, shown as the dashed curves, respectively.
}\label{fidQST}
\end{figure}

   The $N$ spin system is initialized into the input state
$\alpha|\mathbf{0}\rangle+\beta|\mathbf{j}\rangle$ by setting spin
$j$ in the system to  state $\alpha|0\rangle+\beta|1\rangle$. Here
$j$ is the location of the sender (receiver) of the QST for the
(inverse) protocol, which is on some arbitrary spin of the quantum
data bus. It is sufficient to only discuss the transfer of
$|\mathbf{j}\rangle$
because $U_{\tau}$ only introduces a known phase factor before
$|\mathbf{0}\rangle$ [see Eq. (\ref{Ut0})] and $W(c_n,d_n)$ does not
change $|\mathbf{0}\rangle$ [see Eq. (\ref{endgate})]. Iteration $n$
is represented as
\begin{equation}\label{eleO}
Q_{\mathbf{j},n}=[I_{1,...,N-2}\bigotimes
W(c_n,d_n)][U_{\tau}\bigotimes I_{N}].
\end{equation}
After $n$ iterations one obtains
\begin{equation}\label{iterationn}
|\psi_{n}\rangle=T_{\mathbf{j},n}
|\mathbf{j}\rangle=\sum_{k=1}^{N}A_{k,n}|\mathbf{k}\rangle
\end{equation}
using Eqs. (\ref{Utj}-\ref{eleO}). Here
 $T_{\mathbf{j},n}=
 Q_{\mathbf{j},n}...Q_{\mathbf{j},2}Q_{\mathbf{j},1}$,
$A_{N-1,n}=0$, and
\begin{equation}\label{AnP}
A_{N,n}=\sqrt{p_{n}}
\end{equation}
where
\begin{equation}\label{probn}
p_{n}=p_{n-1}+|\langle \mathbf{N-1}|U_{\tau}\bigotimes
I_{N}|\psi_{n-1}\rangle|^2
\end{equation}
with $p_0=0$ and $|\psi_{0}\rangle=|\mathbf{j}\rangle$.
$W(c_{n},d_{n})$ is obtained by setting
\begin{equation}\label{QSTdn}
 d_{n}=e^{i\theta}\sqrt{p_{n-1}}/\sqrt{p_{n}}
\end{equation}
\begin{equation}\label{QSTcn}
   c_{n}=\langle \mathbf{N-1}|U_{\tau}\bigotimes
I_{N}|\psi_{n-1}\rangle/\sqrt{p_{n}}.
\end{equation}
In strict nearest-neighbour chains it can be shown
\cite{BurgarthPRA07} that $p_n$ converges to unity by increasing the
number of iterations. In the present case we have also non-nearest
neighbour interactions, but numerical results show  $p_n$ still
approaches unity, with a convergence speed which depends on the
evolution time $\tau$. [See Figure \ref{fidQST} (a)].
The process of QST after a large number of iterations can be
presented as
\begin{equation}\label{wholeT}
   T_{\mathbf{j},n}(\alpha|\mathbf{0}\rangle+\beta|\mathbf{j}\rangle) \rightarrow
   \alpha
   e^{i n\theta}|\mathbf{0}\rangle+\beta|\mathbf{N}\rangle,
\end{equation}
i.e., spin $N$ ends with the state $\alpha
e^{in\theta}|0\rangle+\beta|1\rangle$ and  $e^{i n \theta}$  is
known.

  We can exploit the inversion of $T_{\mathbf{j},n}$ to implement
the QST from spin $N$ to spin $j$, i.e.,  without applying the
external operation directly on the spin $j$ to evolve it into
state $\alpha|0\rangle+\beta|1\rangle$. Hence the spin chain
functions as a \emph{quantum data bus}, which can transfer
arbitrary unknown states to any qubit. This method also allows to
create a selective excitation that does not require spectral
selectivity, e.g. chemical shift in NMR, to address spin $j$.
 The external
operations are only applied to spins $N-1$ and $N$.
By taking the inner product of Eq. (\ref{iterationn}) with
$\langle\mathbf{N}|$ and using Eq. (\ref{AnP})
one obtains
\begin{equation}\label{fidelityR}
    p_n=|\langle\mathbf{j}|T_{\mathbf{j},n}^{-1}|\mathbf{N}\rangle|^2,
\end{equation}
i.e., $p_n$ is the fidelity for generating $|\mathbf{j}\rangle$ by
applying $T_{\mathbf{n},j}^{-1}$ to $|\mathbf{N}\rangle$. The
creation of the selective excitation for spin $j$  is represented
as
\begin{equation}\label{reverseT}
    T_{\mathbf{j},n}^{-1}(\alpha|\mathbf{0}\rangle+\beta|\mathbf{N}\rangle)\rightarrow
\alpha e^{-i n \theta}|\mathbf{0}\rangle+\beta|\mathbf{j}\rangle.
\end{equation}
By modifying the input state  one can obtain
$T_{\mathbf{j},n}^{-1}(\alpha e^{i n
\theta}|\mathbf{0}\rangle+\beta|\mathbf{N}\rangle)\rightarrow
\alpha |\mathbf{0}\rangle+\beta|\mathbf{j}\rangle$
\cite{footnote1}.

  The method of inverse QST furthermore can be used to entangle arbitrary spins
$j,k$ indirectly by acting at spins $N-1$ and $N$ only.
 This can
be done by designing a pulse analogously to Eq. (\ref{eleO}), and
the required pulse sequence is very similar to the inverse QST.
For this purpose, we set the input state as an entangled state of
a pair of spins $j$ and $k$, represented as
\begin{equation}\label{InputEnt}
|\psi_{jk}\rangle=(|\mathbf{j}\rangle+|\mathbf{k}\rangle)/\sqrt{2}.
\end{equation}
Iteration $n$ can still be represented as Eq. (\ref{eleO}), where
$Q_{\mathbf{j},n}$ is rewritten as $Q_{\mathbf{j,k},n}$, noting
that it depends on the input state. $W(c_n,d_n)$ is obtained in a
similar way by changing $p_0 =
|\langle\mathbf{N}|\psi_{jk}\rangle|^{2}$ and
$|\psi_0\rangle=|\psi_{jk}\rangle$. After a large number of
iterations we obtain
\begin{equation}\label{wholeT2} 
   T_{\mathbf{j},\mathbf{k},n}|\psi_{jk}\rangle \rightarrow
   |\mathbf{N}\rangle.
\end{equation}
where
 $T_{\mathbf{j,k},n}= Q_{\mathbf{j,k},n}...Q_{\mathbf{j,k},2}Q_{\mathbf{j,k},1}$.
 From Eq. (\ref{wholeT2}) one can entangle
spins $j$ and $k$ with high fidelity by applying
$T_{\mathbf{j},\mathbf{k},n}^{-1}$ on $|\mathbf{N}\rangle$,
represented as
\begin{equation}\label{finalent}
T_{\mathbf{j},\mathbf{k},n}^{-1}|\mathbf{N}\rangle \rightarrow
|\psi_{jk}\rangle.
\end{equation}
 The fidelity for generating
$|\psi_{jk}\rangle$ is also represented by Eq. (\ref{fidelityR})
through replacing $T_{\mathbf{j},n}^{-1}$ by
$T_{\mathbf{j},\mathbf{k},n}^{-1}$, and $|\mathbf{j}\rangle$ by
$|\psi_{jk}\rangle$. The numerical simulation for $p_n$ is
illustrated as Figure \ref{fidQST} (b).

\section{Experimental results}
\begin{figure}
\includegraphics[width=4in]{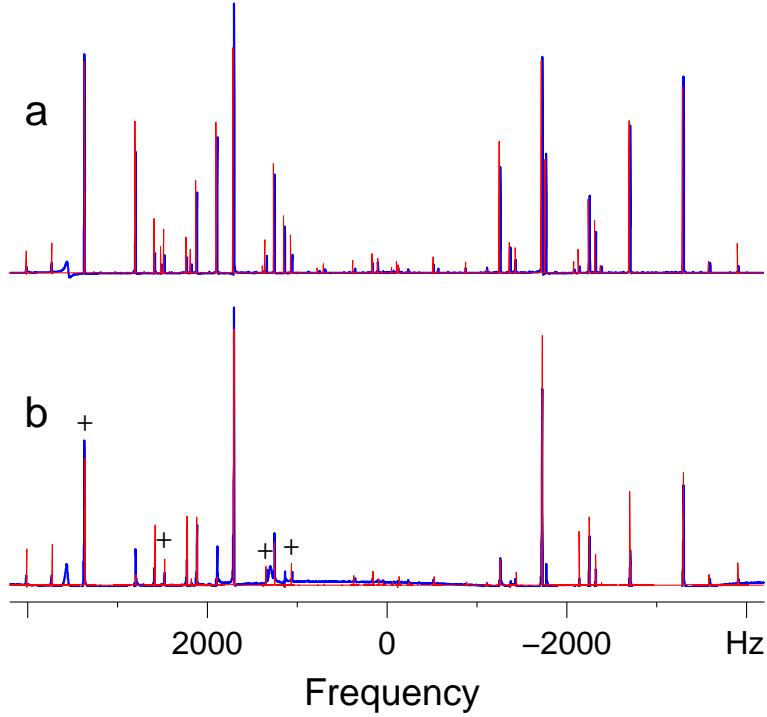}
\caption{(Color online)  NMR spectra  (blue thick) obtained by
Cory48 pulse sequence from the thermal equilibrium state (a), and by
a collective $\pi/2$ pulse from $\rho_{ini}$ (b). The red thin
spectra show the results by simulation. The plot's vertical axes
have arbitrary units. The NMR peaks marked by ''+'' indicate the
single-quantum transitions between magnetic quantum numbers $2$ and
$1$.}
 \label{SpecIni}
\end{figure}

    We use the four protons in ortho- chlorobromobenzene (C$_6$H$_4$ClBr)
dissolved in the liquid crystal solvent ZLI-1132 as four qubits to
implement the experiments.
The Hamiltonian is represented as
\begin{equation}\label{HamNMR}
    H_{NMR} = -\pi\sum_{i=1}^{4}\nu_{i}\sigma_{z}^{i}+\frac{1}{2}\pi
    \sum_{k=2,j<k}^{4}D_{jk}(2\sigma_{z}^{j}\sigma_{z}^{k}-
    \sigma_{x}^{j}\sigma_{x}^{k}-\sigma_{y}^{j}\sigma_{y}^{k}).
\end{equation}
Through fitting the spectra \cite{fittingNMR} obtained by Cory48
\cite{C48ref} and 1-D MREV-8 pulse sequences, and referring to the
spectra of molecules with similar structures
\cite{Mahesh03,cory08LC}, we measure $\nu_1=106.2$, $\nu_2=-187.7$,
$\nu_3=-58.6$, $\nu_4=91.3$ with respect to the transmitter
frequency,
 $D_{12}=-1233.7$, $D_{13}=-149.4$, $D_{14}=-93.2$,
$D_{23}=-716.0$, $D_{24}=-236.6$, $D_{34}=-1677.5$  Hz, and the
effective transverse relaxation times ($T_2^{*}$) as $91$, $87$,
$88$ and $82$ ms \cite{notefit}. The NMR spectrum obtained by Cory48
from the thermal equilibrium state $\rho_{th} =
\sum_{i=1}^{4}\sigma_{z}^{i}$  is shown in Figure
\ref{SpecIni} (a). 

 All experiments start with the
deviation density matrix
   $ \rho_{ini}=|0000\rangle\langle 0000|-|1111\rangle\langle
    1111|$,
which can be prepared by the double-quantum coherence Hamiltonian
\cite{Pine84}
 $H_{d}=\frac{1}{2}\pi
    \sum_{k=2,j<k}^{4}D^{d}_{jk}(\sigma_{x}^{j}\sigma_{x}^{k}-\sigma_{y}^{j}\sigma_{y}^{k})$
in a molecule with C$_{2v}$ symmetry \cite{Furman06}. However, we
choose to generate the effective $H_d$ using a GRadient Ascent
Pulse Engineering pulse \cite{grape}. Using temporal averaging, we
prepare $\rho_{ini}$ by summing the three states
$U_d\rho_{th}U_d^{\dag}$, $U_d^{\dag}\rho_{th}U_d$, $2\rho_{th}$,
 where $U_d=e^{-i t_d H_d}$ by choosing
$t_d=8.00/D_{12}^{d}$ \cite{Furman06}. 
In the numerical simulation we
prepare $\rho_{ini}$ with fidelity $99.97\%$.
 Figure \ref{SpecIni} (b) shows
the NMR spectrum obtained by a collective $\pi/2$ pulse in the
experiment when the system lies in $\rho_{ini}$. The NMR peaks
marked by ''+'' indicate the single-quantum transitions between
magnetic quantum numbers $2$ and $1$.

\begin{figure}
\includegraphics[width=4in]{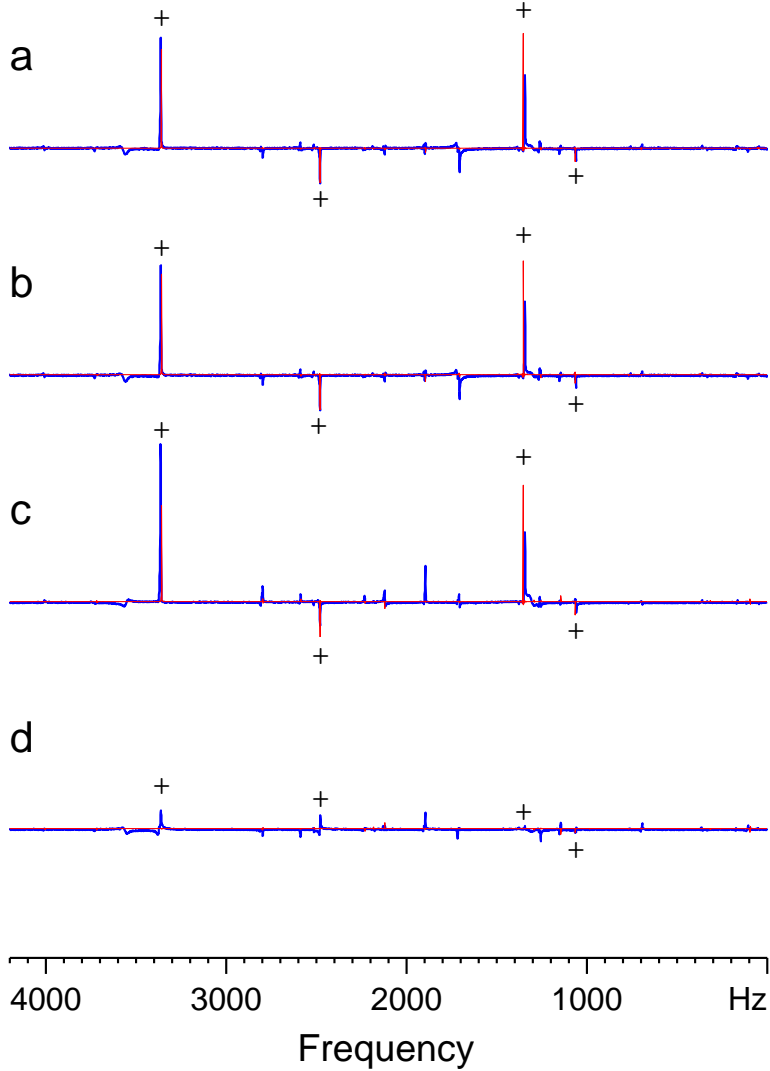}
\caption{(Color online)
NMR spectra (a-d) for implementing the QST from spins $1$ to $4$
after $100$ iterations, when the input states are chosen as
$\sigma_{x}|000\rangle\langle000|$,
$\sigma_{y}|000\rangle\langle000|$,
$\sigma_{z}|000\rangle\langle000|$ and $I|000\rangle\langle000|$
respectively, where the readout operation $e^{i\pi\sigma_{y}^{4}/4}$
is applied to obtain observable signals in (c) or (d). The plot's
vertical axes have the same scale.
}
 \label{Spec}
\end{figure}

  We demonstrate the QST by transferring $\rho_{0}$  from spins $1$ to $4$
by choosing  $\rho_{0}=\sigma_{x}$, $\sigma_{y}$,
  $\sigma_{z}$ and $I$, respectively.
Because $T_{\mathbf{j},n}$ is spin-preserving, the transitions
marked by ''+'' in Figure \ref{SpecIni} (b) can represent the QST
starting with the input state $\rho_0|000\rangle\langle000|$. We
therefore can ignore $|1111\rangle\langle1111|$ in $\rho_{ini}$ and
omit the negative frequency spectral region.

  The input state is prepared by
applying an operation $U_{ini}$
to $\rho_{ini}$. With increasing $n$, $T_{\mathbf{1},n}$ transforms
$\rho_{0}|000\rangle\langle000|$ to $|000\rangle\langle000|\rho$
asymptotically, where $\rho=e^{i n\theta\sigma_{z}/2}\rho_{0}e^{-i
n\theta\sigma_{z}/2}$. In experiments we removed the phase factor
between $\rho$ and $\rho_0$ by phase correction. For a fixed $n$, we
implement the unitary $T_{\mathbf{1},n}U_{ini}$ using one GRAPE
pulse. The experimental results of the QST after $100$ iterations
for the various input states are shown as Figures \ref{Spec} (a-d),
respectively. 

  Exploiting the transformation between the computational basis and energy eigenbasis,
   and ignoring the difference of $T_{2}^{*}$ of the four protons, we can
approximatively obtain $A_{k,n}$ in Eq. (\ref{iterationn}) through
measuring the amplitudes of the peaks marked by "+" in Figures
\ref{Spec} (a-c) by choosing the signals in Figure \ref{SpecIni} (b)
as the reference. Therefore we obtain $p_n=|A_{4,n}|^{2}$. For the
input states $\sigma_{x}|000\rangle\langle000|$,
$\sigma_{y}|000\rangle\langle000|$,
$\sigma_{z}|000\rangle\langle000|$, $p_{100}$ is measured as
$0.654\pm 0.046$, $0.660\pm0.052$ and $0.693\pm0.037$, respectively.
All other
$|A_{k,100}|^{2}$ are below $0.02$. 

  To observe $p_n$ increasing with $n$, we also implement the QST by
choosing various $n$ when the input state is
$\sigma_{x}|000\rangle\langle000|$. The measured $p_n$ is shown in
Figure \ref{fidQST} (a) as the data marked by ''$\times$'', which
can be fitted as $0.65p_n$.
\begin{figure}
\includegraphics[width=4in]{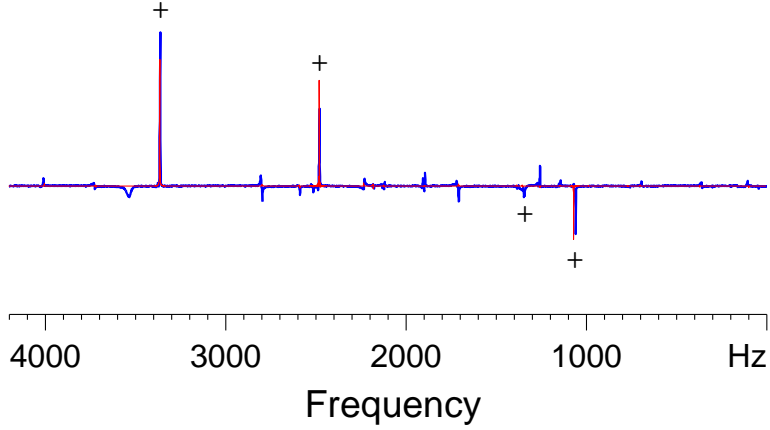}
\caption{(Color online) NMR spectra for implementing the selective
excitation and quantum data bus for spin $2$. }
 \label{selective}
\end{figure}

  Next we implement the selective excitation / quantum data bus for spin $2$.
The reverse QST starts with the input state
   $ |000\rangle\langle000|\sigma_{x}$
obtained by applying $R_{y}^{4}=e^{-i\sigma_{y}^{4}\pi/4}$ to
$\rho_{ini}$. When $n=100$, $T_{\mathbf{2},n}^{-1}$ transforms
$|000\rangle\langle000|\sigma_{x}$
  to
$|0\rangle\langle0|\rho|00\rangle\langle00|$ with probability
close to $1$, where $\rho=e^{-i n\theta\sigma_{z}/2}\sigma_{x}e^{i
n\theta\sigma_{z}/2}$. The experimental results are shown in
Figure \ref{selective}. The fidelity of excitation is measured as
$0.744\pm0.036$.

\begin{figure}
\includegraphics[width=4in]{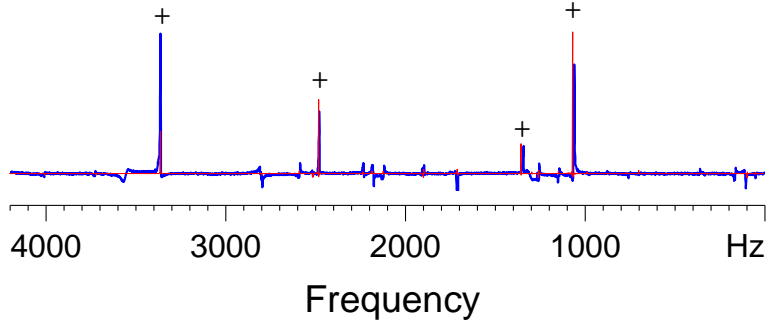}
\caption{(Color online) NMR spectra for measuring the fidelity of
the generation of
$(|\mathbf{1}\rangle+|\mathbf{4}\rangle)/\sqrt{2}$. }
 \label{LDEfig}
\end{figure}

 We choose
$|\psi_{14}\rangle=(|\mathbf{1}\rangle+|\mathbf{4}\rangle)/\sqrt{2}$
as the target to demonstrate the entangling operation in spins $1$
and $4$. To measure the fidelity, we rewrite Eq. (\ref{fidelityR})
as $p_{n}=|\langle0000|\Psi_n\rangle|^2$ \cite{zhang09} by
replacing $|\mathbf{j}\rangle$ by $|\psi_{14}\rangle$. Here
$|\Psi_n\rangle=P^{\dag}T_{\mathbf{1},\mathbf{4},n}^{-1}|\mathbf{4}\rangle$
where
$P$ denotes the operation to prepare $|\psi_{14}\rangle$ from
$|0000\rangle$ (e.g. see \cite{Chuang98}). When $p_n$ is close to
$1$, we can obtain $p_n$ approximately by applying a readout
operation $e^{i\pi\sigma_{y}^{1}/4}$ to $|\Psi_n\rangle$. Noting
that $|1111\rangle\langle1111|$ in $\rho_{ini}$ does not
contribute observable signals for measuring $p_n$, we approximate
$p_n$ as the coherence
\begin{equation}\label{Coherence}
  C_n=|Tr(2|0000\rangle\langle1000|\rho_n)|
\end{equation}
where
   $ \rho_n=U_{tot,n}
    \rho_{ini}U_{tot,n}^{\dagger}$
with
$U_{tot,n}=e^{i\pi\sigma_{y}^{1}/4}P^{\dagger}T_{\mathbf{1},\mathbf{4},n}^{-1}e^{i\pi\sigma_{x}^{4}/2}$.
The simulated and measured $C_n$ is shown in Figure \ref{fidQST}
(b).  
The experimental data can be fitted as $0.77C_n$. Figure
\ref{LDEfig} illustrates the NMR spectra when $n=8$.

The operations  $U_d$, $U_d^{\dag}$, $R_{y}^{4}$,
$T_{\mathbf{1},n}U_{ini}$, $T_{\mathbf{2},n}^{-1}R_{y}^{4}$, and
$U_{tot,n}$ are experimentally implemented using the GRAPE pulses
with fidelities in theory larger than $0.99$, respectively. The
pulse lengths are $10$ ms for $U_d$ and $U_d^{\dag}$, $20$ ms for
the other pulses. The experimental errors could mainly result from
the inhomogeneities of the magnetic field, imperfect
implementation of GRAPE pulses and decoherence. In order to
estimate the quality of the experimental spectra, we also list the
ideal ones in simulation, shown as the red thin curves in Figures
\ref{SpecIni}-\ref{LDEfig}.

\section{conclusion}
   We have given an NMR implementation for various important
tasks of quantum control that in principle can be achieved {\em
indirectly} by controlling the end of a spin chain. The dipolar
couplings naturally existing in the liquid crystal NMR system are
directly exploited for the QST. The experimental results
demonstrate the successful control of the spin system with dipolar
couplings by the GRAPE pulses. Firstly, we implemented the
transfer of an arbitrary quantum state.  Secondly, by implementing
the reverse QST, we have created a full quantum data bus which is
controlled by the two-qubit end gates. Finally as another
application of the reverse QST, we proposed and demonstrated a new
method to implement an entangling operation.

\section{Acknowledgments}
We thank D. Cory, C. Ramanathan, S. Bose, G. B. Furman and T. S.
Mahesh for helpful discussions. We acknowledges support by the
EPSRC grant EP/F043678/1, NSERC, QuantumWorks, Premier Discovery
Award and CIFAR.


\end{document}